\documentstyle[12pt]{article}

\begin{document}

\title{Quasi-Local ``Conserved Quantities''}

\author{Roh S. Tung\\
 {\it California Institute for Physics and Astrophysics}\\
 {\it 366 Cambridge Avenue, Palo Alto, California 94306, USA}\\
 {\it E-mail: tung@calphysics.org}\\
 {\it and}\\
 {\it Enrico Fermi Institute, University of Chicago}\\
 {\it 5640 South Ellis Avenue, Chicago, Illinois 60637, USA}}
  \maketitle

\begin{abstract}
Using the Noether Charge formulation, we study a perturbation of
the conserved gravitating system. By requiring the boundary term
in the variation of the Hamiltonian to depend only on the
symplectic structure, we propose a general prescription for
defining quasi-local ``conserved quantities'' (i.e. in the
situation when the gravitating system has a non-vanishing energy
flux). Applications include energy-momentum and angular momentum
at spatial and null infinity, asymptotically anti-deSitter
spacetimes, and thermodynamics of the isolated horizons.\\
\\
\\
Proceedings of the 9th Marcel Grossmann Meeting, World Scientific.
 \end{abstract}

\newpage

Consider a general diffeomorphism-invariant field theory in $n$
dimensions with a Lagrangian $n$-form ${\cal L}(\phi)$, where
$\phi$ denotes an arbitrary collection of dynamical fields. The
equations of motion, ${\cal E}=0$, are obtained by computing the
first variation of the Lagrangian,
\begin{equation}
\delta {\cal L}={\cal E} \delta \phi  + d \Theta (\phi ,
\delta\phi ),
\end{equation}
where $\Theta (\phi,\delta\phi)$ is the symplectic potential
$(n-1)$-form\cite{Wald1993}. For any diffeomorphism generated by a
smooth vector field $\xi^a$ , one can define a conserved Noether
current $(n-1)$-form $J(\xi)$ (also known as the ``covariant
Hamiltonian $(n-1)$-form'' \cite{CNT1995} in the first order
formulation) by
\begin{equation}
J(\xi)=\Theta(\phi,{\rm\pounds}_\xi\phi)-i_\xi {\cal L},
\end{equation}
where ${\rm\pounds}_\xi$ denotes the Lie derivative and $i_\xi$ is
the inner product. On shell, the Noether current is closed, and
can be written in terms of an $(n-2)$-form $Q(\xi)$ (the Noether
charge) as $J(\xi)=d Q(\xi)$. The variation of the Noether current
$(n-1)$-form is given by
\begin{equation}
\delta J(\xi)= \omega(\phi,\delta\phi,{\rm\pounds}_\xi\phi) + d
i_\xi \Theta(\phi,\delta\phi) ,
\end{equation}
where $\omega$ is the symplectic current $(n-1)$-form defined by
\begin{equation}
\omega(\delta_1\phi,\delta_2\phi)=\delta_1\Theta(\phi,\delta_2\phi)-
\delta_2\Theta(\phi,\delta_1\phi) .
\end{equation}
Its integral over a Cauchy surface $\Sigma$ defines the
presymplectic form $\Omega$ . Note that it has ambiguities
inherent in the definition, but the ambiguities (in $\Theta$) do
not appear to give rise to an ambiguity in the definition of
$\Omega$ for suitable asymptotic conditions on the dynamical
fields. We assume that such suitable asymptotic conditions have
been imposed so that the presymplectic form is well-defined
\cite{Wald1994} .

Now, consider a perturbation process from an initial state at
$\Sigma_0$ to a final state at $\Sigma$ such that $\xi^a$  is
normal to both $\partial\Sigma_0$  and  $\partial\Sigma$.
Integrate and vary the perturbed Noether current $(n-1)$-form
$\Delta J(\xi)=J(\xi)\vert_{\partial\Sigma}
-J(\xi)\vert_{\partial\Sigma_0}$ we obtain
\begin{equation}
\delta \int \Delta J(\xi)=\int \Delta
\omega(\phi,\delta\phi,{\rm\pounds}_\xi\phi)+\oint i_\xi
\Delta\Theta(\phi,\delta\phi) .
\end{equation}
In order for the boundary to have a symplectic structure, we
subtract a total variation term,
 $$
\delta \int \Delta J(\xi)-d i_\xi\Theta(\phi,\Delta\phi) =\int
\Delta \omega(\phi,\delta\phi,{\rm\pounds}_\xi\phi)+\oint i_\xi
[\Delta\Theta(\phi,\delta\phi)-\delta\Theta(\phi,\Delta\phi)] ,
 $$ 
then the last term has a symplectic structure. Now we are ready to
define a ``conserved quantity''  ${\cal H}(\xi)$ by
\begin{equation}
{\cal H}(\xi)= \int \Delta J(\xi)-d i_\xi\Theta(\phi,\Delta\phi),
\end{equation}
and thus its variation is
\begin{equation}\label{variationH}
\delta {\cal H}(\xi) =\int \Delta
\omega(\phi,\delta\phi,{\rm\pounds}_\xi\phi)- \oint i_\xi
\omega(\phi,\delta\phi,\Delta\phi) .
\end{equation}
In conclusion we briefly consider applications in the following
two cases:
\begin{itemize}

\item
Case I, $\oint i_\xi \omega(\phi,\delta\phi,\Delta\phi)=0$.

In this case, there exist a Hamiltonian  and is given by
$H(\xi)={\cal H}(\xi)$. Examples are (a) spatial infinity: by
taking the initial state $\phi_0$ to be flat spacetimes, it gives
the ADM mass \cite{ReggeTeitelboim1974} ;  (b) asymptotically
anti-deSitter spacetimes: by taking the initial state $\phi_0$ to
be the anti-deSitter metric, we obtain the desirable conserved
quantities \cite{AshtekarDas2000} ; (c) For black hole event
horizons or isolated horizons\cite{Ashtekar2000}, take the vector
field $\xi^a$ to be the killing field or the null vector normal to
the horizons. In these cases, conserved quantities are
well-defined.

\item
Case II, $\oint i_\xi \omega(\phi,\delta\phi,\Delta\phi)\ne 0$.

In this case, there does not exist a Hamiltonian but ${\cal
H}(\xi)$ defines a ``conserved quantity''. It satisfies the
consistency check\cite{WaldZoupas2000} ($\delta^2 {\cal
H}(\xi)=0$), and since it depends only on the symplectic
structure, ${\cal H}(\xi)$ is unique (independent of which
dynamical variables are chosen) and well-defined (although it is
not conserved). Our proposed definition (\ref{variationH})
includes a nonzero energy flux such that ${\rm\pounds}_\xi {\cal
H}(\xi)=-\oint i_\xi\omega(\phi,{\rm\pounds}_\xi\phi,\Delta\phi)$
gives the energy loss formula. This arises in radiating systems
(such as at null infinity) as described by Wald and
Zoupas\cite{WaldZoupas2000}. At null infinity, it gives the Bondi
energy formula. By considering the boundary at the isolated
horizons, it is possible that we may use the ideas presented here
to study perturbations of the thermodynamical quantities such as
entropy.

\end{itemize}

We would like to thank Jim Nester and Bob Wald for helpful
discussions.


\end{document}